\documentclass[manuscript]{acmart}

\usepackage{amsmath,amsfonts}
\usepackage{algorithmic}
\usepackage{algorithm}
\usepackage{array}
\usepackage[caption=false,font=normalsize,labelfont=sf,textfont=sf]{subfig}
\usepackage{textcomp}
\usepackage{stfloats}
\usepackage{url}
\usepackage{verbatim}
\usepackage{tikz}
\usetikzlibrary{arrows.meta, shapes, positioning, calc}
\usepackage{graphicx}
\usepackage{subcaption}
\usepackage{xcolor}
\usepackage{comment}
\usepackage{tabularx}
\usepackage{booktabs}
\usepackage{fontawesome}
\usepackage{bm}
\usepackage{xspace}
\usepackage{float}
\usepackage{accents}
\usepackage{threeparttable}
\usepackage{diagbox}
\usepackage{textcomp}
\usepackage[skins]{tcolorbox}
\usepackage{enumitem}
\usepackage{multirow}
\usepackage{longtable}
\usepackage{hyperref}
\usepackage{caption}
\usepackage{ulem}
\usepackage[switch]{lineno}
\usepackage{ragged2e}
\usepackage{utfsym}

\def\BibTeX{{\rm B\kern-.05em{\sc i\kern-.025em b}\kern-.08em
    T\kern-.1667em\lower.7ex\hbox{E}\kern-.125emX}}

\newcommand{\Code}[1]{\begin{small}\fontsize{9.5}{10}\selectfont\texttt{#1}\end{small}}
\newcommand{\SmallCode}[1]{\begin{scriptsize}\texttt{#1}\end{scriptsize}}

\newlength{\fsize}
\newcommand{\myindent}{\hspace*{2em}}

\tcbset{
  my box/.style={
    enhanced,
    colframe=#1!80,
    colback=#1!10,
    attach boxed title to top left={xshift=0.2cm, yshift=-0.2cm},
    boxed title style={
      colback=#1!80,
      outer arc=0pt,
      arc=0pt,
      top=0pt,
      bottom=0pt,
    },
  },
}
\newtcolorbox{result-rq}[1]{
  my box=black,
  title=#1,
  boxrule=1.2pt,top=6pt,bottom=3.5pt,left=6pt,right=6pt
}

\begin{document}

\title{Analyzing C/C++ Library Migrations at the Package-level: Prevalence, Domains, Targets and Rationals across Seven Package Management Tools}

\author{Haiqiao Gu}
\affiliation{%
 \institution{Peking University}
 \city{Beijing}
 \country{China}}

\author{Yiliang Zhao}
\affiliation{%
 \institution{Peking University}
 \city{Beijing}
 \country{China}}

\author{Kai Gao}
\affiliation{%
 \institution{University of Science and Technology Beijing}
 \city{Beijing}
 \country{China}}

\author{Minghui Zhou}
\affiliation{%
 \institution{Peking University}
 \city{Beijing}
 \country{China}}

\begin{abstract}
Library migration, i.e., replacing a third-party library with another, happens when a library can not meet the project's requirements and is non-trivial to accomplish. 
To mitigate the problem, substantial efforts have been devoted to understanding its characteristics and recommending alternative libraries, especially for programming language (PL) 
ecosystems with a central package hosting platform, such as Python (PyPI), JavaScript (npm), and Java (Maven). 
However, to the best of our knowledge, understanding of C/C++ library migrations is still lacking, possibly due to challenges resulting from the fragmented and complicated dependency management practices in the C/C++ ecosystem. Given the widespread use and critical role of C/C++, 
the lack of such understanding hinders the formulation of best practices and tools to facilitate library migrations in the C/C++ ecosystem.

To bridge this knowledge gap, this paper analyzes 124,786 dependency configuration file changes in 19,943 C/C++ projects that utilize different package management tools and establishes the first C/C++ library migration dataset, comprising 2,191 migrations and 717 migration rules. Based on the dataset, we investigate the prevalence, domains, target library, and rationale of C/C++ library migrations and compare the results with three widely investigated PLs: Python, JavaScript, and Java. 
We find that the overall trend in the number of C/C++ library migrations is similar to Java, which grew slowly and stabilized, and then decreased slightly. Notably, the number of migrations in emerging package management tools, such as Meson and Vcpkg, is on the rise. Migrations across different package management tools are also observed, with the most common being from Git submodule to Conan.
In C/C++, library migrations mainly occur in three domains, i.e., GUI, Build, and OS development, but are rare in domains (e.g., Testing and Logging) that dominate library migrations in the three compared PLs. 
83.46\% of C/C++ source libraries only have one migration target, suggesting that our library migration dataset could be used directly to recommend migration targets for C/C++ developers. 
We find four C/C++-specific migration reasons, such as less compile time and unification of dependency management, revealing the unique dependency management requirements in C/C++ projects. 
We believe our findings can help C/C++ developers make more informed library migration decisions and shed light on the design of C/C++ library migration tools.
The scripts and dataset can be accessed at \url{https://github.com/Dl0qWmJ4/C-CPPLibraryMigration}
\end{abstract}

\maketitle




\section{Introduction}
\label{sec:intro}

Current software projects commonly reuse third-party libraries to boost productivity and ensure code quality~\cite{DBLP:npm_trivial_pkgs,DBLP:java_pkg_usage,DBLP:java_test_usage}. 
However, reusing third-party libraries is not a one-time solution. As the project and third-party libraries evolve, the reused libraries may not meet the project's functional, legal, or performance requirements~\cite{he_large-scale_2021}, and may even introduce security vulnerabilities~\cite{DBLP:npm_vulner, DBLP:npm_vul}. When such issues arise, a commonly adopted solution by developers is to replace the reused library with another one featuring the same or similar functionality, referred to as \textit{library migration}.

Library migration is a time-consuming and effort-intensive task that involves target library selection and code adaptation~\cite{DBLP:python_test_migration, DBLP:java_method_mig, DBLP:lib_selection_discuss}. 
To ease the task, substantial efforts have been devoted to understanding the characteristics of library migrations~\cite{teyton2012mining,teyton2014study,he_large-scale_2021, DBLP:npm_vulnerable_migration, gu_self-admitted_2023, DBLP:python_test_migration, DBLP:pymigbench, python_library_migration_2024} and recommending target libraries~\cite{DBLP:npm_package_recommend, he_migrationadvisor_2021, he_multi-metric_2021} and APIs~\cite{DBLP:java_method_mig, DBLP:java_api_recommend, DBLP:android_api_migration, DBLP:api_semantic_mining}. These studies primarily focus on PL ecosystems with central package hosting platforms, such as Maven for Java, npm for JavaScript, and PyPI for Python, where dependent libraries can be easily and uniquely identified from unified and structural dependency configuration files. 


However, similar topics are rarely explored in the C/C++ ecosystem, while C/C++ developers also encounter challenges with finding alternatives during migration\cite{ExampleOfMigration1, ExampleOfMigration2, ExampleOfMigration3}
(below is an example from stackoverflow).
\begin{quote}
    \textit{In our project now we using log4cxx, but those library don't develop some years, also we have some problems with it. Could you advise some library for logging in C++. Library must support multithread logging, system-log ...
    Also lib license must be very democracy ...
    Must support linux, windows. 
    }
    \cite{ExampleOfMigration1}
\end{quote}

Unlike other PL ecosystems, dependency management in the C/C++ ecosystem is notoriously fragmented and complex~\cite{miranda2018use}. 
Developers often rely on a mix of \textit{package management tools} (PMTs), such as Conan and Git submodules, along with custom build systems and manual configurations~\cite{DBLP:tang22}.
When migrations occur, they may need to maintain multiple dependency configuration files.
And the huge number and differences among packages hosted on various package management platforms further pose challenges for developers seeking alternatives.
%
Compiling C/C++ packages across various platforms also burdens them. 
These characteristics make C/C++ developers particularly conservative when introducing or modifying dependencies, which may result in significant differences in library migration practices compared to other PLs. 
Given the popularity and significant role of C/C++ in critical areas such as operating systems (OS), embedded systems, graphic user interface (GUI), and high-performance computing, we argue that it is urgent to advance the understanding of library migrations in the C/C++ ecosystem, especially its unique characteristics compared to those in other ecosystems. Such an understanding could provide developers, maintainers, and stakeholders with valuable information to make more informed decisions and design automated tools for C/C++ library migrations. 

Therefore, we attempt to bridge the gap in this study by analyzing a large body of C/C++ migrations. Specifically, we retrieve 717 migration rules and 2,191 migration instances from 124,786 dependency configuration file changes across 19,943 C/C++ projects. Based on this dataset, we analyze the prevalence, domains, target library, and rationale of C/C++ library migrations and compare the results with three widely studied PLs, i.e., Python, JavaScript, and Java. 
In particular, we answer the following research questions (RQs):
\begin{itemize}
    \item \textbf{RQ1:} \textit{How common do C/C++ migrations occur?} 
    \\
    \textit{Motivation}: 
    This question aims to understand the prevalence and trend of C/C++ migrations and the migration frequency across different package management tools, which can highlight the urgency of investigating C/C++ library migrations and facilitate library migrations for specific package management tools in a more targeted way.\\
    \textit{Findings}: The number of C/C++ library migrations grew slowly, stabilized gradually, and then decreased slightly, which is similar to that in the Java ecosystem. 
    We find that the number of library migrations in emerging tools such as Meson and Vcpkg undergoes rapid growth. 
    We also observe that some developers choose to switch between different package management tools, with the most common PMT migration pattern being from Git submodule to Conan.
    \item \textbf{RQ2:} \textit{In which domains do C/C++ migrations occur?} \\
    \textit{Motivation}: Migration domains vary across different ecosystems. This question aims to understand the most common domain for C/C++ in which migrations occur and reveal frequent migration domains that require more attention.\\
    \textit{Findings}: Build and GUI dominate nearly half of C/C++ library migrations.
    Notably, there are relatively few C/C++ migrations in domains like logging and testing (utility libraries), which take the most library migrations of the three compared PL ecosystems. 
    \item \textbf{RQ3:} \textit{What is the distribution of C/C++ migration targets chosen by developers?}\\
    \textit{Motivation}: 
    The migration targets represent the technology trend or best practices that developers turn to, and their distribution reflects the divergence of technical trends in the C/C++ ecosystem. 
    This question attempts to characterize the migration alternatives in the C/C++ ecosystem, providing insights for effectively selecting migration targets.\\
    \textit{Findings}: Source libraries in most (83.46\%) C/C++ library migration rules have only one migration target. 
    C/C++ library migrations exhibit considerably stronger one-to-one and unidirectional characteristics than those in other PLs, indicating that universal migration practices may have been formed in C/C++. This finding also suggests that mining more comprehensive and accurate migration rules may be the right direction to recommend migration targets. 
    \item \textbf{RQ4:} \textit{Why do developers conduct C/C++ migrations?} \\
    \textit{Motivation}: This question aims to understand the rationale for migrations in C/C++ and to find developers' specific considerations for C/C++ migration.
    \\
    \textit{Findings}: We uncover four C/C++ specific migration reasons, such as less compile time and unification of dependency management. 
    The main cause of C/C++ migrations includes the functionality of the target library and integration with platforms/other dependencies. 
\end{itemize}

Our findings provide unique insights into the evolution of package management tools, migration recommendations, and library selection, to guide and inform developers on effectively managing dependencies and migrations in C/C++ projects.
We publicize our C/C++ migration dataset for further investigation. 

\begin{figure*}[t]
    \centering
    \includegraphics[width=\linewidth]{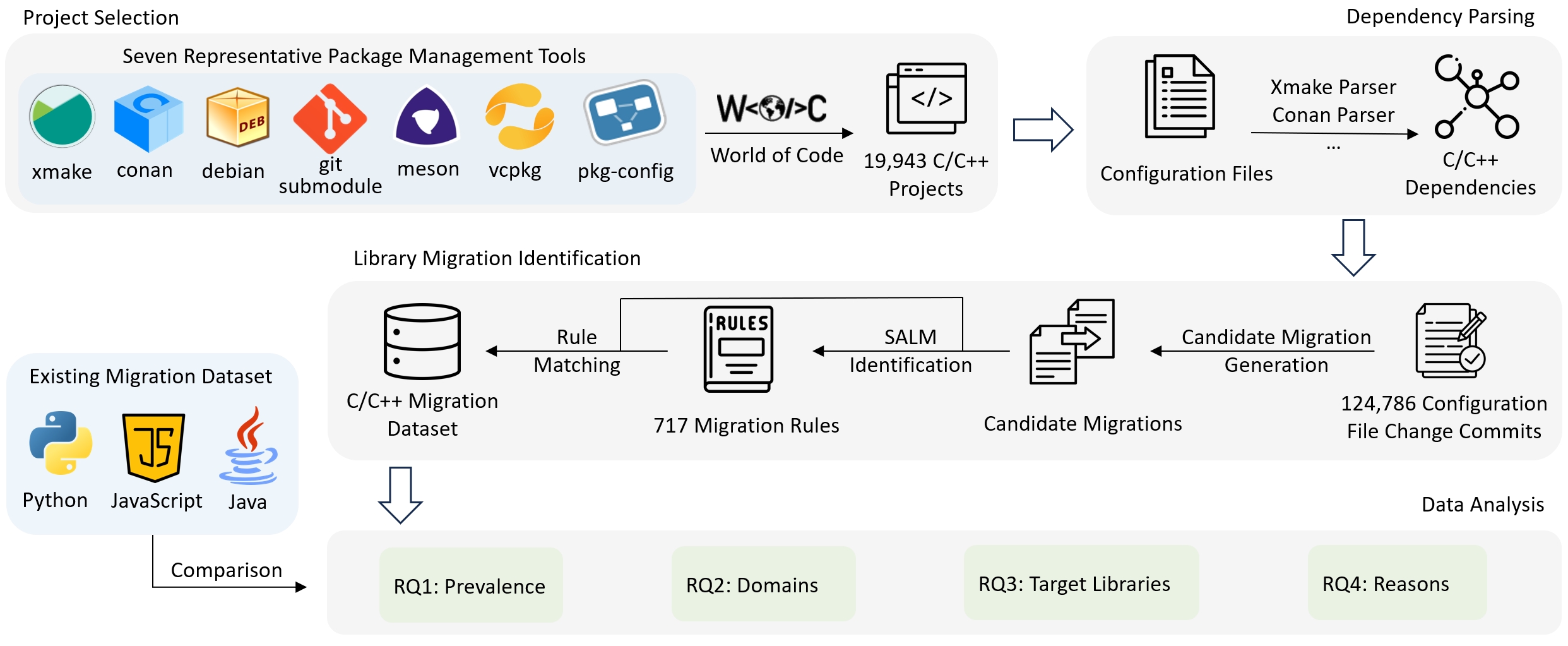}
    \caption{Overview of our study}
    \vspace{-3mm}
    \label{fig:methodology}
\end{figure*}
\section{Background and Related Work}

Library migration refers to the process of replacing one library or dependency (source library, $l_1$) with another library 
\footnote{library with another name. For C/C++, the name of the library may contain version number, resulting in different names for different versions of the library (libmutter-2, libmutter-3). We regard this kind of change as a library update rather than a migration}
(target library, $l_2$) that offers similar functionality\cite{kabinna2016logging, he_multi-metric_2021, teyton2012mining, teyton2014study}. 
Notably, due to developers mixing multiple package management approaches in C/C++, a library may be removed from one package management tool while being added to another.  
In such cases, the modified package may have the same name. We also consider this a kind of library migration, i.e., PMT migration, which has not been discussed in previous studies.
A migration rule is denoted as $\langle l_1, l_2\rangle$ in previous studies\cite{he_multi-metric_2021,teyton2014study,he_large-scale_2021,gu_self-admitted_2023}. A PMT migration rule is denoted as $\langle pmt_1, pmt_2\rangle$.

Migration is often necessary to address various issues, such as deprecated features\cite{mujahid_where_2023}, security vulnerabilities, performance improvements\cite{alrubaye2020does,kula2018developers}, or compatibility requirements with other components of the software project\cite{Xu_2023}. It can also be driven by the need to take advantage of new features\cite{kula2018developers}, better support\cite{barbosa2022and}, or licensing considerations\cite{Xu_2023} provided by alternative libraries\cite{gu_self-admitted_2023,he_large-scale_2021}.

\textbf{Empirical studies have been conducted to understand the nature of library migrations.} 
Teyton et al.\cite{teyton2012mining, teyton2014study} mined 1,198 migrations from 15,168 Java projects and summarized eight migration reasons.
Kabinna et al.\cite{kabinna2016logging} identified 49 logging library migrations in 223 Apache Software Foundation projects and found that flexibility and performance are the primary drivers. 
He et al.\cite{he_large-scale_2021} analyzed the prevalence, trends, and rationales of the library migration phenomenon occurring in the Java ecosystem on a large-scale dataset (3,163 migration commits from 19,652 projects). They discovered four dominant domains (logging, JSON, testing, and web service) and 14 reasons for Java project migrations.
Gu et al.\cite{gu_self-admitted_2023} compared self-admitted library migrations (SALMs) occurring in the Java, JavaScript, and Python ecosystems.

\textbf{Various methods have also been proposed to recommend alternative libraries}, leveraging the historical migration data from software repositories. These methods tap into the collective wisdom of developers' decisions to migrate to alternative packages~\cite{teyton2012mining,teyton2014study,DBLP:java_api_mig_tool}. 
However, they face challenges such as low recall or low precision in identifying suitable replacements. Taking it a step further, He et al.\cite{he_multi-metric_2021} transformed the recommendation problem into a ranking problem and devised some more subtle metrics to improve the efficiency and accuracy of recommendations. 
To avoid undesired suggestions, Mujahid et al.\cite{mujahid_where_2023} proposed an automated approach to identify npm packages in decline and suggest superior alternatives.

Although there has been significant research on library migration in PLs that have central package hosting platforms (e.g., Maven for Java\cite{alrubaye2020does,teyton2014study,he_large-scale_2021,he_migrationadvisor_2021,he_multi-metric_2021,gu_self-admitted_2023,teyton2012mining, DBLP:java_api_mig_tool}, npm for JavaScript\cite{mujahid_where_2023,barbosa2022and,gu_self-admitted_2023}, and PyPI for Python\cite{gu_self-admitted_2023,kabinna2016logging,islamcharacterizing}) and considerable datasets on these migrations exist\cite{alrubaye2020does,he_large-scale_2021,gu_self-admitted_2023}, the C/C++ ecosystem has received comparatively less attention. 
In this study, we define the \textbf{C/C++ ecosystem} as software projects primarily using C/C++ as the main language and their dependent libraries.
To the best of our knowledge, there is no study investigating the landscape and rationale of library migration that occurs in the C/C++ ecosystem. No specific tool or methodology for migration mining and recommendation is currently available in the C/C++ ecosystem, either. Partially inspired by similar work on other languages, in this study, with a rule-based migration mining algorithm, we construct the first large-scale C/C++ migration dataset across seven package management tools  
to conduct an empirical study on the prevalence, domain, targets, and rationale of migrations in the C/C++ ecosystem.

\section{Approach}

\label{sec:approach}

To construct a dataset that can reflect the landscape of C/C++ library migrations, we first select C/C++ projects that adopt certain package management tools~(Section~\ref{ss:dataselection}). 
Then, we parse the dependencies of these projects utilizing specific parsers~(Section~\ref{ss:dependencyparsing}).
After that, we identify SALM rules based on the change of dependencies. Finally, we apply the rules to the entire history of the change to construct the final migration dataset for the C/C++ ecosystem.~(Section~\ref{ss:identify_mig})
The approaches we used for subsequent data analysis are described in each research question, respectively.
Figure~\ref{fig:methodology} presents the logic flow of the approach adopted in this study.

\subsection{Project Selection} \label{ss:dataselection}
A dataset covering all the C/C++ projects is ideal for exploring our research questions. 
To achieve a comprehensive data scope, we begin with \textit{World of Code} (WoC), which is a large-scale open source dataset and platform\footnote{https://bitbucket.org/swsc/overview/src/master/}. WoC 
provides comprehensive and unified access to almost all public Git repositories on dozens of platforms, including GitHub, GitLab, Bitbucket, and so on. 
It organizes data based on the git data structure (blob, tree, and commit) and stores mappings of other kinds of data like authors, projects, files, and so on.
We can easily obtain the data we need via WoC.

Given the breadth of PMTs identified by Tang et al.\cite{DBLP:tang22}, our study narrowed its focus to seven PMTs, as shown in Table~\ref{tab:files}, for three reasons. 
First, they are widely used in C/C++ projects~\cite{DBLP:tang22} and cover different types of PMTs. For example, Deb is widely adopted by projects in the operating system domain, which constitutes a key and unique application domain of C/C++, and whose importance and popularity differ from those of other programming languages. Conan, Vcpkg, and Xmake reflect emerging PMTs in the C/C++ ecosystem aiming to provide package hosting platforms. 
Second, they adopt structural file formats such as JSON and config as configuration files, which allows us to parse dependencies with a high accuracy. 
Third, other PMTs are either rarely used or it is difficult to accurately parse dependencies specified in them, such as CMake and Autoconf, which threatens the validity of the results. 
In total, we obtain 19,943 C/C++ projects $\mathcal{P}$ that use any of the seven PMTs from WoC.

\subsection{Dependency Parsing}
\label{ss:dependencyparsing}
We design a parser for each dependency configuration file adopted by the seven PMTs. The details of each parser are listed in Table~\ref{tab:files}.
We use different parsing approaches to resolve specific dependency fields for different PMTs. For example, we use regular expressions to parse the \Code{dependency} field of \Code{meson.build} (\Code{dependency('libplacebo', version: '>= 3.110.0', required: false)}), and just match declaration patterns to parse \Code{Requires} field of \Code{*pc} (\Code{Requires: actionlib\_msgs std\_msgs trajectory\_msgs})

For validation, ten different configuration files are sampled for each PMT. Compared to previous work with only 15 samples~\cite{DBLP:tang22}, our 70 samples are quite sufficient to assure the accuracy of our result.
Two authors with three years of C/C++ development experience independently checked all seventy files and labeled the dependencies introduced by certain PMTs. We use Krippendorff’s alpha with MASI distance~\cite{krippendorff2018content} to measure the inter-rater agreement of dependencies and get a result of 0.96. After discussion, all disagreements are resolved and the label result serves as our ground truth.
Our dependency parsing method obtains a precision of 84\% and a recall of 90\% on the ground truth. 
Most false positives are caused by multiple choices in if-statements or local projects built by developers. Most false negatives are caused by the declaration of dependencies in script files(\Code{.py/.lua}). 
However, these false positives will be excluded in our following approach to identifying SALMs by matching migrated libraries in commit messages.
Therefore, this parsing method and results are sufficient for our subsequent analysis needs.

\label{sec:parse}
\begin{table}[t]
    \centering
    \caption{Configuration Files Parsed in Our Method}
    \label{tab:files}
    \vspace{-1mm}
    \renewcommand{\arraystretch}{1.2}
    \footnotesize
    \begin{tabular}{>{\centering\arraybackslash}m{1.4cm}>{\centering\arraybackslash}m{1.9cm}>{\centering\arraybackslash}m{1.4cm}>{\centering\arraybackslash}m{2.1cm}}
        \toprule
            \textbf{PMTs} & \textbf{Configuration Files} & \textbf{Parsing Approach} & \textbf{Dependency Field(s)} \\
        \midrule
        Meson & meson.build & regex & \SmallCode{dependency} \\
        Xmake & xmake.lua & regex &  \SmallCode{add\_requires} \\
        Deb & debian/control & regex & \SmallCode{build-depends}, \SmallCode{depends} \\
        Conan & conanfile.txt/py & regex/AST & \SmallCode{build\_requires}, \SmallCode{requires} \\
        Vcpkg & vcpkg.json & JSON parser & \SmallCode{dependencies}  \\
        Pkg-config & *.pc & pattern match & \SmallCode{Requires}\\
        Git submodule & .gitmodules & pattern match & \SmallCode{submodule} \\
        \bottomrule
    \end{tabular}
    \vspace{-1mm}
\end{table}

\subsection{Library Migration Identification}~\label{ss:identify_mig}
To obtain a comprehensive dataset containing all the possible C/C++ library migrations as accurately as possible, we go through three phases. First, we extract dependency changes in selected projects to generate candidate migrations.
Second, we identify \textit{self-admitted library migrations (SALMs)}~\cite{gu_self-admitted_2023} from these candidates to achieve migration rules; third, we utilize the migration rules to rematch the candidate migrations we have collected, thereby obtaining a more comprehensive migration dataset.

\subsubsection{Candidate Migration Generation}
\label{ss:candidategeneration}
Commits to C/C++ configuration files suggest changes in C/C++ configuration, indicating possible changes on dependencies, i.e., possible migrations.
We utilize the WoC dataset, specifically version V, compiled in March 2023~\cite{ma2019world}, to mine commits $\mathcal{C}$ related to selected C/C++ configuration files in projects $\mathcal{P}$, along with their pre- and post-modification blobs. 

Through static analysis mentioned in Section~\ref{sec:parse}, we parse the blob contents before and after each commit to identify the introduction or removal of third-party C/C++ libraries.
For each commit $c \in \mathcal{C}$, developers may introduce new libraries ($L^+_c$), remove old libraries ($L^-_c$), or both. All these involved libraries constitute the set of libraries $\mathcal{L} = \bigcup_{c\in \mathcal{C}} L^+_c$ (Note that $\bigcup_{c\in \mathcal{C}} L^-_c\subseteq\bigcup_{c\in \mathcal{C}} L^+_c$ ).
We consider only commits $\mathcal{C}_{cand}$ with both library introduction and removal (\textbf{candidate commits}), i.e., $\mathcal{C}_{cand}=\{c\ |\ c \in \mathcal{C}\wedge L^-_c\neq\phi\wedge L^+_c\neq\phi\}$.
Then, we generate pairs of libraries that are removed and introduced for each $c$ as \textbf{candidate migrations} (denote as $\mathcal{M}_{cand} = \{\langle c, l_1, l_2 \rangle\ |\ c \in \mathcal{C}_{cand} \wedge l_1\in L^-_c \wedge l_2\in L^+_c\}$), suggesting potential library migrations.
4,922,903 candidate migrations are generated in this step, corresponding to 45,133 candidate migration commits.
\subsubsection{SALM-based Migration Rule Identification and Filter}
\label{ss:ruleidentification}
A candidate migration $m_{cand}$ is considered as a migration $m$ only when it is confirmed by \textbf{migration rules}.
We followed previous work\cite{gu_self-admitted_2023} to identify migration rules based on self-admitted library migrations (SALMs). In SALMs, developers explicitly document migration behavior in commit records so that accurate migration rules can be obtained automatically by matching library names. For PMT migrations, we replace the match of library names with the match of PMT names. 
However, the existing migration mining algorithms suffer from certain limitations in our study.

First, the naming conventions of different PMTs pose challenges in matching C/C++ library names in commit messages.
Each PMT has its own rule and name strategy. For different PMTs, we adopt corresponding strategies to ensure the accuracy of migration identification(e.g., drop \Code{lib-}, \Code{-dev}, or extract library name from git URL).

Second, similar package names can lead to numerous incorrect migration pairs. Gu et al. partially mitigated this issue by aggregating similar packages into super libraries. However, this approach needs a lot of manual effort and results in the loss of detailed information about individual packages. In our work, we significantly reduced these false positives by matching the most similar package names, thereby further improving the accuracy and automation of migration identification.

We further eliminated library update cases from the migration results.
For example, C/C++ libraries in Deb often have the version number appended to the library name during release (e.g., \Code{libmutter-2}, \Code{libmutter-3}). Candidate migration involving only version number changes should be identified as library updates rather than migrations.
However, we reserve major version updates similar to those from qt4 to qt5, which is a complicated transition from one collection of libraries to another (e.g., \Code{libqt4-dev} to \Code{qtbase5-dev}, \Code{qttools5-dev}, and \Code{qttools5-dev-tools}).
In Gitsubmodule, developers may also change another git URL with different owner names yet the same repository name (i.e., a fork project).
We detect these candidate migrations between fork or direct version number changes by rules and exclude them as trivial migrations.
The candidate migration from \Code{debhelper} to \Code{debhelper-compat} in Deb is also excluded because this change is a constraint on versions of \Code{debhelper} instead of migration between build-related libraries.
717 migration rules $\mathcal{R}$ are identified by SALMs in this step.

Due to the lack of publicly available datasets for C/C++ migration, the validation of our mining algorithm is conducted through a manual review process based on the relevance of commit messages and the similarity of involved libraries. 
251 rules out of 717 are sampled (0.95 confidence level and 0.05 margin of error), and two authors manually checked them independently, resulting in a Krippendorff's alpha of 0.87. A 96\% precision is obtained after resolving all disagreements.

\subsubsection{Migration Dataset Construction}
\label{ss:datasetconstruction}
Migration rules $\mathcal{R}$ identified from SALMs are used to confirm whether candidate migrations $\mathcal{M}_{cand}$ correspond to migrations $\mathcal{M}$, i.e., $\mathcal{M} = \{\langle c, l_1,l_2\rangle\ |\ \exists \langle l_1, l_2\rangle\in \mathcal{R}, \langle c, l_1,l_2\rangle\in\mathcal{M}_{cand}\}$. 
Similarly, we denote migration commits as $\mathcal{C}_m = \{c\ |\ \exists \langle c, l_1,l_2\rangle\in\mathcal{M}, c\in\mathcal{C}_{cand}\}$.
This approach is based on the inference that a commit has also conducted a migration if a migration rule occurs in its dependency changes, even if developers may not document it in its records.
Through applying migration rules to candidate migrations, 2,191 migrations ($\mathcal{M}$, corresponding to 1,737 commits) are obtained and serve as the dataset for our subsequent analysis.
The distribution of migrations among seven PMTs is shown in Table~\ref{tab:mig-dataset}.

\begin{table}[t]
\centering
\renewcommand{\arraystretch}{1.5}
\caption{An overview of our dataset}
\footnotesize
\label{tab:mig-dataset}
    \begin{tabular}{lrrrrrr}
        \toprule
        \textbf{PMTs} & $|\bm{\mathcal{P}}|$ & $|\bm{\mathcal{P}_m}|$ & $|\bm{\mathcal{C}_{cand}}|$ & $|\bm{\mathcal{C}_m}|$ & $|\bm{\mathcal{M}}|$ & $|\bm{\mathcal{R}}|$  \\ \midrule
        Conan        & 552    & 133   & 2,200   & 224   & 266   & 104  \\ 
        Vcpkg        & 241    & 65   & 2,703   & 92   & 96   & 53 \\ 
        Meson        & 500  & 156    & 1,960   & 253   & 279   & 105 \\ 
        Xmake        & 121    & 4     & 364     & 5     & 6     & 5  \\ 
        Pkg-config   & 29     & 0     & 81      & 0     & 0     & 0 \\ 
        Gitsubmodule & 11,067 & 107   & 18,927  & 113   & 123   & 86 \\ 
        Deb          & 7,433 & 868 & 18,968  & 1,050 & 1,421 & 364 \\ \midrule
        Total        & 19,943 & 1,333 & 45,133 & 1,737 & 2,191 & 717 \\ \bottomrule
    \end{tabular}
\end{table}

\section{RQ1: How common do C/C++ migrations occur?}

The complex nature of dependency management of C++ brings uncertainty to the nature of library migration. In this question, we want to know the prevalence of migrations in C/C++, compare its differences with other languages, and get deeper into the differences among different PMTs. 

\subsection{RQ1.1 How common do C/C++ migrations occur in different package management tools?}
\subsubsection{Method}

Based on our dataset, we calculate both the number of migrations and projects with at least one migration to   the lower bound of the number of migrations.
For projects, the set of projects that have ever conducted migrations is denoted as $\mathcal{P}_m = \{p\ |\ p\in\mathcal{P}\wedge \exists c\in C_p, c\in\mathcal{C}_m\}$ if we let $C_p$ be the set of commits for $p$. We compare the number of $\mathcal{C}_m$, $\mathcal{M}$, and $\mathcal{P}_m$ among the seven PMTs. Finally, we plot the trend of overall migrations $\mathcal{M}$ and migrations in each PMT by commit timestamps.

\subsubsection{Results}

As shown in Table~\ref{tab:mig-dataset}, 6.68\% of projects (1,333~/~19,943) have conducted at least one migration within the studied scope, and 3.85\% of candidate commits (1,737~/~45,133) are migration commits.
The number of migrations differs vastly across PMTs.
Deb contributes 65.11\% of migrations (868~/~1,333), which is far more than those in other PMTs because of its long history and wide usage.
Conversely, only 123 migrations are identified in Git submodule, despite its history spanning over a decade, where 31.32\% of candidate commits (5,927~/~18,927) are just changing another fork repository and thus excluded from this study.
Xmake presents the fewest migrations (only six), and we identified no migrations in Pkg-config.
That may be because only a few projects adopting these two PMTs are analyzed, or developers did not explicitly document the migration in commit messages.

\begin{figure}
    \centering
    \includegraphics[width=0.7\columnwidth]{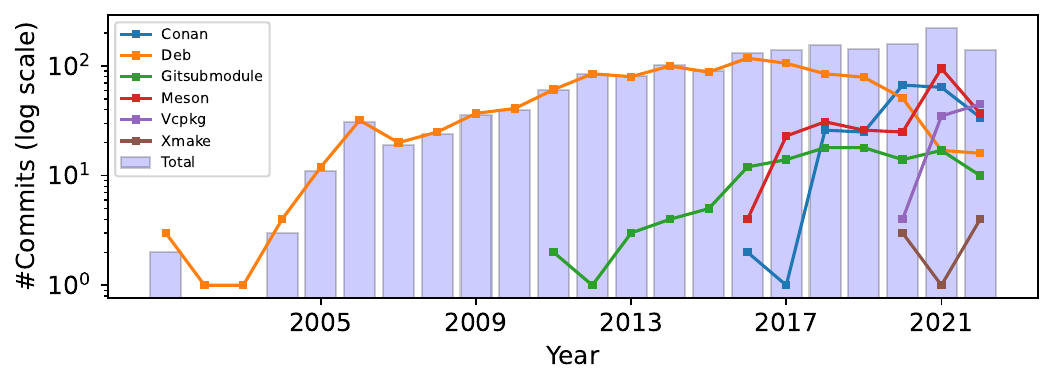}
    \vspace{-5mm}
    \caption{Number of migration commits in C/C++}
    \label{fig:commit_freq}
    \vspace{-3mm}
\end{figure}

To better understand the migration trend in C/C++ PMTs, we plot the longitudinal trend of migration commits (histogram in Fig.~\ref{fig:commit_freq}) and projects with $\geq 1$ migrations (not shown because the trend is highly similar to the number of migration commits) in log scale,
which starts in 2001 and ends in 2023, because the data trim in March 2023. 
We find that both the total number of migration commits and projects increased slowly (until 2016) and finally remained relatively stable.
The lines in Fig.~\ref{fig:commit_freq} with different colors demonstrate the trend of migration commits across different PMTs.
We find that although the overall trend has remained relatively stable since 2016, the amount of migrations fluctuates significantly across different PMTs. 
It is worth noting that migration commits in Meson sharply increased in 2020 (when \textit{Meson 0.54.0} was released, adding support for the CMake project and library), quickly surpassing the rapidly declining Deb, making Meson the PMT with the most migrations.

\textit{\textbf{Comparison with other PLs:}}

As a mature programming language, the trend of migrations in C/C++ PMTs is similar to Java~\cite{gu_self-admitted_2023}. They both experienced a gradual increase and then a stabilizing trend. 
On the one hand, this stabilized migration trend reflects the maturity of both C/C++ and Java ecosystems, where major libraries are well-established. In newer languages like Python and JavaScript, libraries are often subjected to significant overhauls, introducing new features and deprecating old ones at a faster pace. The stable feature of C/C++ libraries facilitates maintenance and long-term planning, which also enhances the feasibility of automated migration tools.

On the other hand, despite the long development history of C/C++, its dependency management has been continuously evolving mostly since it doesn't have a universal package hosting platform (in other words, it has fragmented package management tools). 
To tackle the problem, new tools have emerged, e.g., Meson, Conan, and Vcpkg. 
Migrations in these PMTs keep growing because libraries under them are evolving rapidly, leading to a more dynamic migration landscape, which behaves similarly to Python. 

\vspace{-1mm}
\begin{result-rq}{Summary for RQ1.1:}
Among the 19,943 projects, 1,333 (6.68\%) have conducted at least one migration. The overall amount of migration commits in C/C++ increased slowly and remains stable with a slightly decreasing trend, the same as Java. 
However, the number of migrations 
and projects that conducted migrations 
in emerging package management tools (Meson, Vcpkg) is increasing, while the migration in those established (Deb), although initially higher, is experiencing a decline. 

\end{result-rq}

\subsection{RQ1.2: How common do C/C++ migrations occur across different package management tools?}

\subsubsection{Method}

Due to the lack of a universal way for C/C++ dependency management, developers may also \textbf{migrate their PMTs} during the development of projects. In this part, we calculate this migration with a similar identification approach as SALMs, replacing `libraries` with `PMTs`. 306 PMT migrations (291 confirmed manually) are identified, and we drop the duplicated results that occur in a single project. 141 projects with PMT migration commits and 15 PMT migration rules are identified. 15 of them mentioned why they changed their PMTs. This type of migration is significantly different from the conventional migrations typically discussed; therefore, here we only discuss its frequency of occurrence and underlying causes.

\subsubsection{Results}

\begin{figure}
    \centering
    \includegraphics[width=0.7\columnwidth]{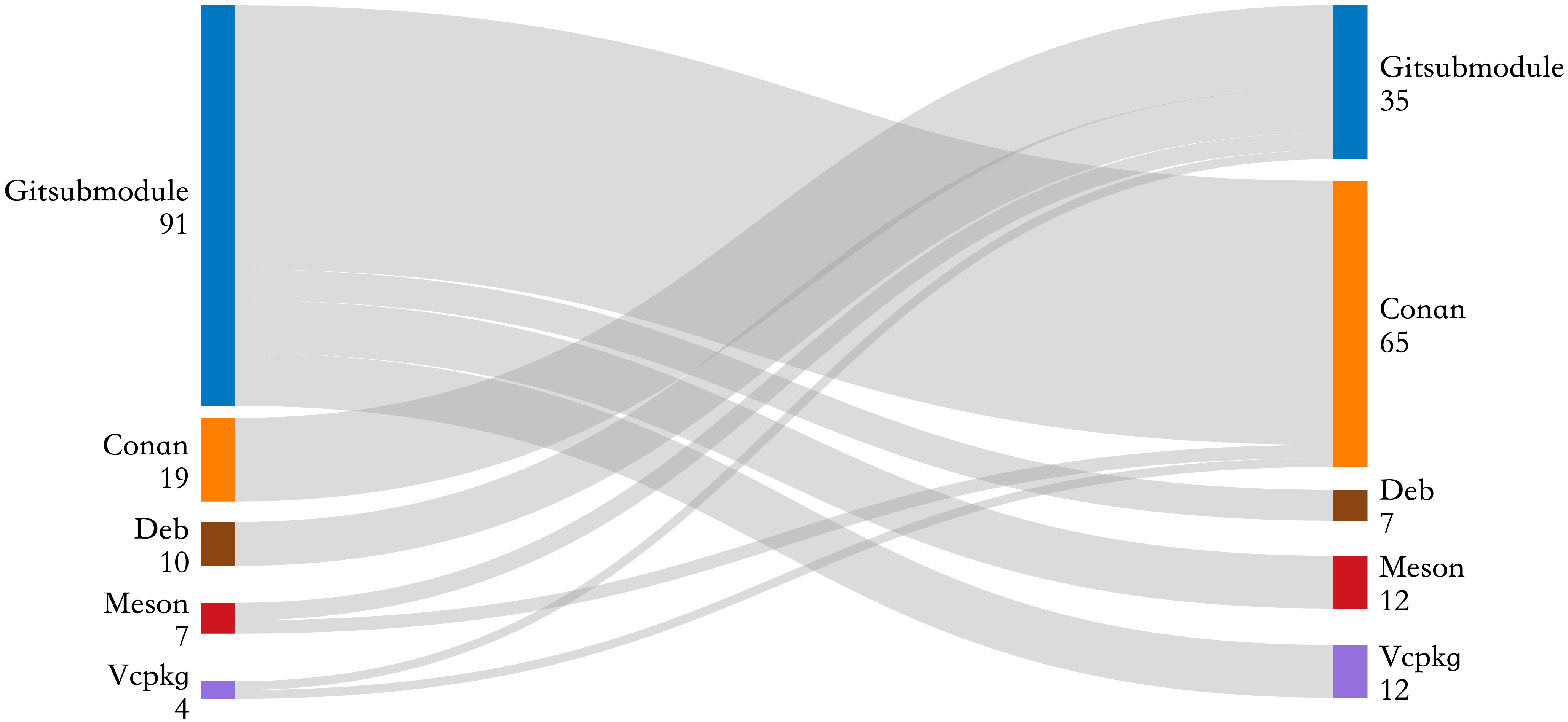}
    \vspace{-5mm}
    \caption{PMT migrations in C/C++}
    \label{fig:cross_sankey}
    \vspace{-3mm}
\end{figure}

Fig.~\ref{fig:cross_sankey} shows the PMT migration rules that occur more than once. The most frequent migration rule is from Git submodule to Conan (60 projects). 47.5\% (67 / 141) of projects switch submodules to different package managers to mitigate the effort of dependency management. Opposite migrations also exist, such as 19 projects changing Conan to Git submodule. The reason for such a phenomenon may be two major characteristics of the C/C++ ecosystem: one is that each PMT employs a unique management strategy, resulting in varying levels of compatibility and usability; another is the widespread practice of code cloning in C/C++. Using Git submodules, developers can maintain their git branches or keep synced with the latest snapshots.

\vspace{-1mm}
\begin{result-rq}{Summary for RQ1.2:}
One hundred and forty one (141)
C/C++ projects are observed to have changed their package management tools (compared to 1,333 projects with migrations). 
While migrations from Git submodule to Conan are most frequent, 
 the opposite migration path is also observed.

\end{result-rq}

\section{RQ2: In which domains do C/C++ migrations occur?}

Migration domains vary across different ecosystems,
reflecting the application of PL, as well as its evolution in ecology. 
This question aims to understand the most
common domain for C/C++ in which migrations occur
and reveal frequent migration domains that require more
attention.

\subsection{Method}

Given the decentralization of C/C++ libraries, our analysis covers all 416 libraries related to migrations that we have identified. 
We retrieve descriptions or README.md of these libraries from respective sources to classify them into different categories.
Libraries from Conan/Vcpkg are obtained from certain hosting platforms (conan.io, github.com/vcpkg).
Libraries from meson/xmake/gitsubmodule are obtained from GitHub or download URLs.
Libraries from Deb are obtained from the Debian package hosting platform.
362 of 416 libraries are obtained in this way and the remaining libraries are obtained through manual search.
Then, we randomly sample 100 libraries from the 416 libraries. One author labels them with different domains and generates a coding book. Another author uses this coding book to again label the 100 randomly selected libraries, resulting in a Krippendorff's Alpha of 0.84. 
Second, after resolving the disagreements and ensuring clear and accurate expression in the coding book, two authors label the remaining libraries independently.
Finally, we get 19 distinct domains in total. 
These domains cover common results as other languages (e.g., network, database, testing) and C/C++ specific domains (e.g., OS development). 
Fig.~\ref{fig:domain_dist} presents the domain distribution of migrations. Domains with fewer than 20 migrations are merged and displayed as 'Other' in the figure. We also visualize three sub-graphs for the top three domains using Sankey diagrams~\cite{DBLP:sankey_diagram} to show how migrations flow in these domains.

\subsection{Results}


\begin{figure}[t]
    \centering
    
    \begin{minipage}[b]{0.49\columnwidth}
        \centering
        \includegraphics[width=\linewidth]{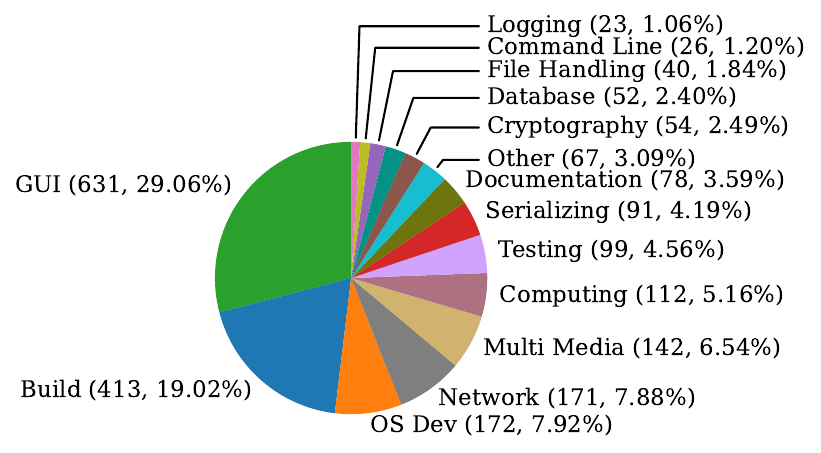}
        \caption{Distribution of domains for migrations in C/C++}
        \label{fig:domain_dist}
    \end{minipage}
    \hfill
    \begin{minipage}[b]{0.49\columnwidth}
        \centering
        \includegraphics[width=\linewidth]{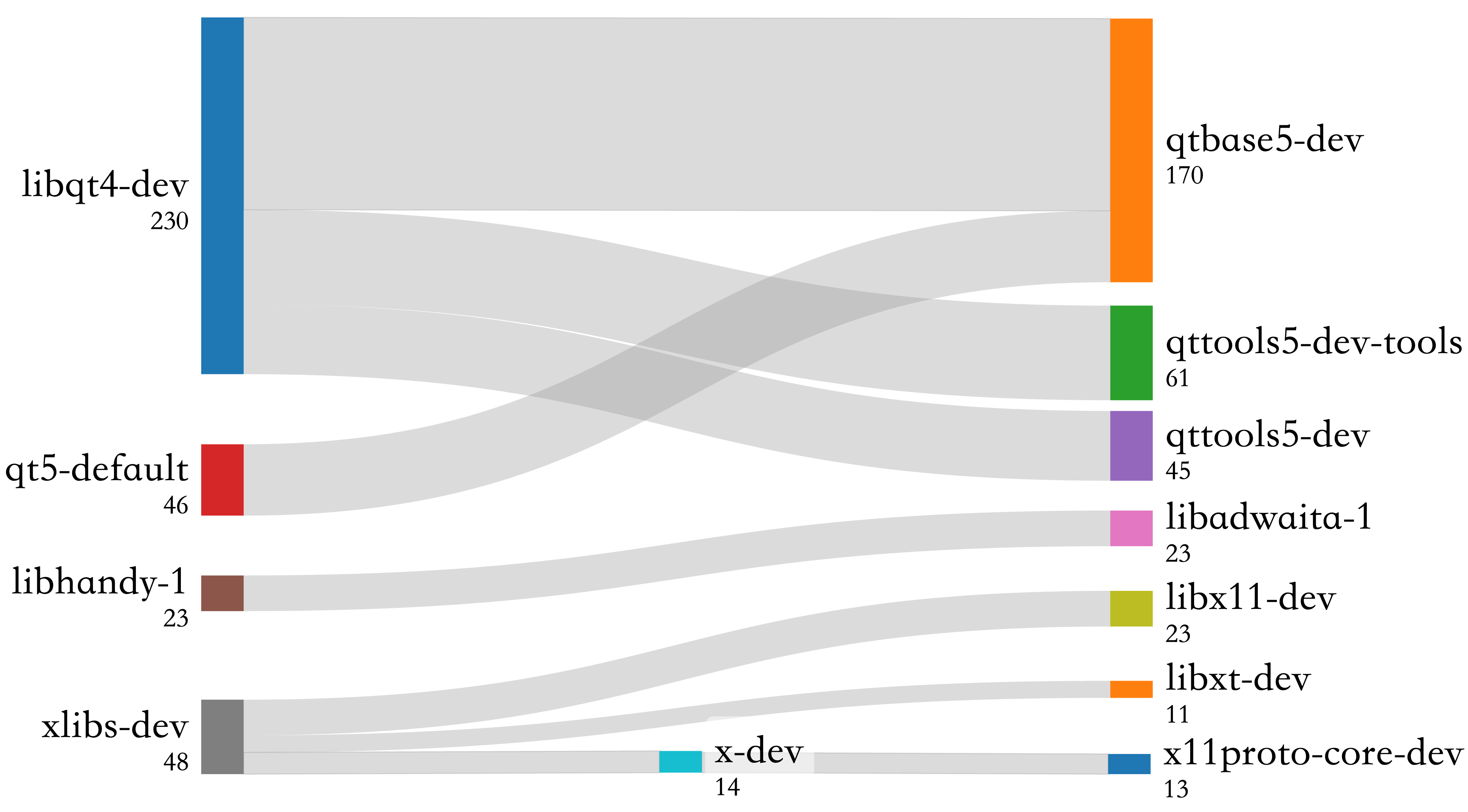}
        \caption{Migration subgraph for GUI libraries}
        \label{fig:sub1}
    \end{minipage}
    \hfill
    \\
    \vspace{2mm}
    \begin{minipage}[b]{0.49\columnwidth}
        \centering
        \includegraphics[width=\linewidth]{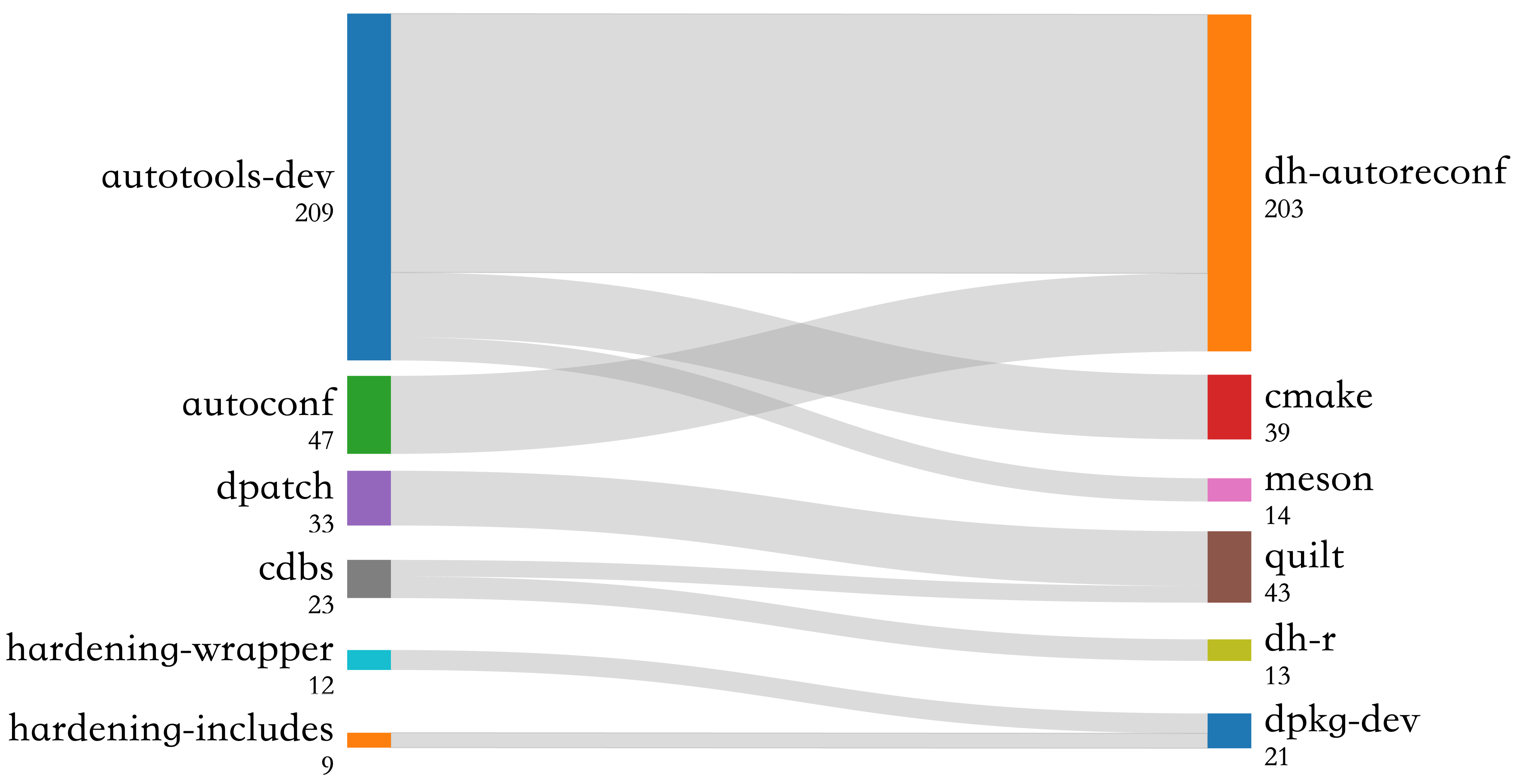}
        \caption{Migration subgraph for Build libraries}
        \label{fig:sub2}
    \end{minipage}
    \hfill
    \begin{minipage}[b]{0.49\columnwidth}
        \centering
        \includegraphics[width=\linewidth]{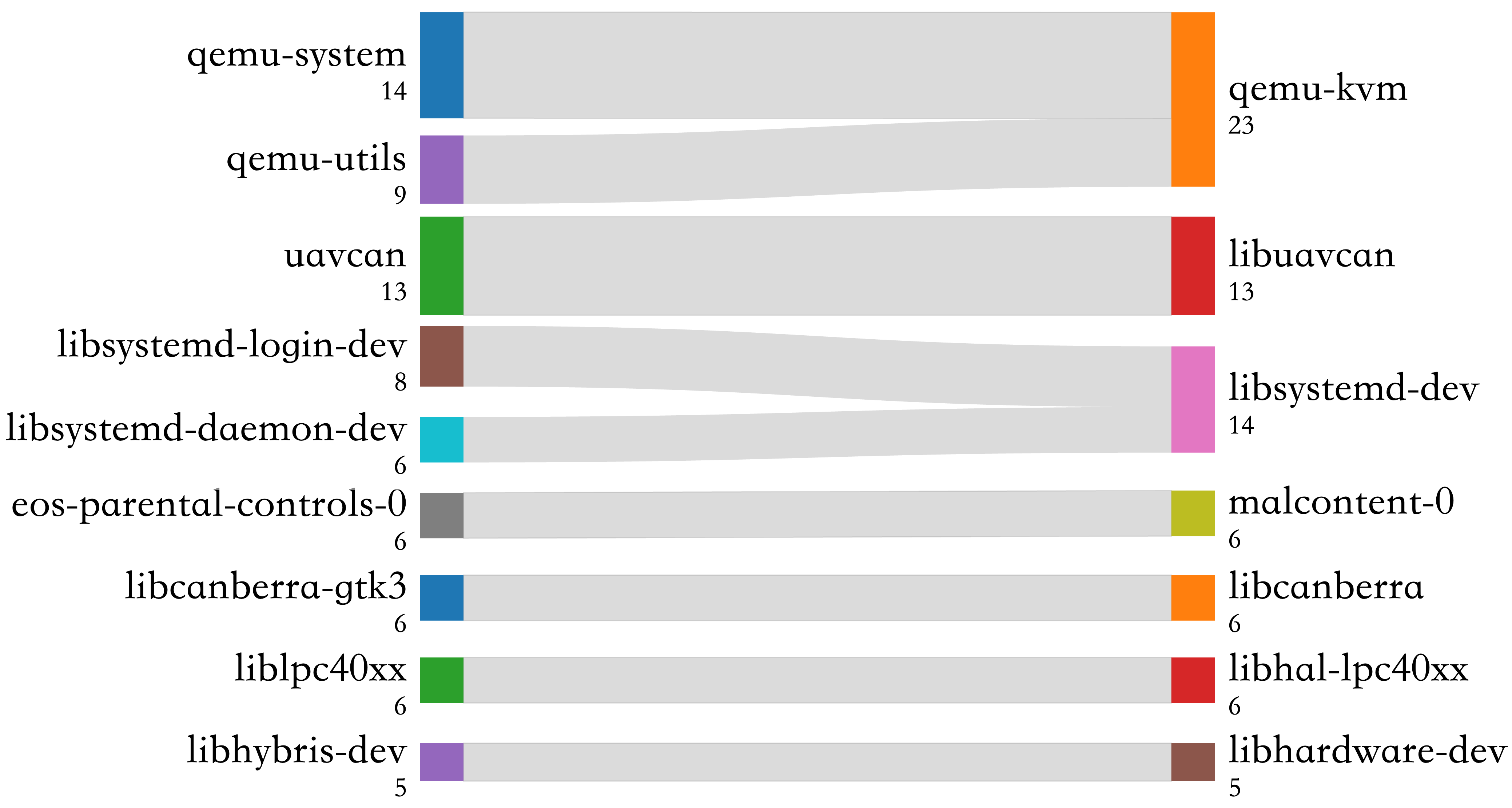}
        \caption{Migration subgraph for OS Development libraries}
        \label{fig:sub3}
    \end{minipage}
    
    \label{fig:three_subfigures}
\end{figure}

Two domains (Build and GUI) occupy nearly half of the migrations (48.08\%), followed by OS development, which is also an essential application domain of C/C++, accounting for 7.92\%.
However, Testing and Logging, which are critical migration areas in other ecosystems, only account for 4.56\% and 1.06\%, respectively. 
The details of top three domains with the most migrations are as follows:

\textbf{GUI}: GUI occupies the highest proportion (29.06\%) in C/C++ migrations.
In this domain, migrations may be primarily driven by technological innovations.
The most frequent migration within this domain is from Qt4 to Qt5 (Fig.~\ref{fig:sub1}).
The "upgrade" from Qt4 to Qt5 is accompanied by significant performance optimization and usage updates. It led to the fragmentation of the original package (\textit{libqt4-dev}) into multiple components (\textit{qtbase5-dev}, \textit{qttools5-dev}, \textit{qttools5-dev-tools}), which is why we regard this as a kind of migration.
Another typical migration is from \textit{libhandy-1} to \textit{libadwaita-1}.
The two libraries aimed at providing adaptive, responsive UI widgets for GNOME applications, while libhandy-1 is rooted in the GTK3 era, and libadwaita-1 represents the modern, GTK4-based direction of development. 

\textbf{Build}: Build libraries refer to the libraries or tools used during packaging or distributing. Their dominance of the migration domain in the C/C++ ecosystem may be due to the complexity of the build process of Debian packages. The most frequent migrations are $\langle \textit{autotools-dev,\ dh-autoreconf} \rangle$ (156 migrations). 
\textit{dh-autoreconf} was introduced to automate the regeneration of \textit{Autotools} configuration files during the Debian package build process. The Debian packaging system is a mature and widely adopted method for managing operating system libraries, and it also handles numerous C/C++ libraries.
By eliminating the need for static helper files provided by \textit{autotools-dev}, it simplifies maintenance and enhances cross-platform compatibility, which led to its gradual migration to \textit{autotools-dev}. Other migrations in this domain ($\langle \textit{autoreconf,\ dh-autoreconf} \rangle$, $\langle \textit{dpatch,\ quilt} \rangle$) also show the essence of functionality and ease of maintenance for the build process.

\textbf{OS Development}: 
In this domain, developers migrate libraries on process/memory/session management, file system, or drivers. Migrations in this domain may be the result of specific functional needs for platforms and architectures.
Frequent migrations are $\langle \textit{libsystemd-login-dev,\ libsystemd-dev} \rangle$ and $\langle \textit{qemu-system,\ qemu-kvm} \rangle$. The first migration resembles a consolidation of functionalities, while the second is tailored for another runtime environment. 
There are also migrations involving libraries across different interaction layers with hardware ($\langle \textit{liblpc40xx,\ libhal-lpc40xx} \rangle$), where \textit{liblpc40xx} provides direct hardware-level APIs, while \textit{libhal-lpc40xx} leverages those low-level functions to offer a more accessible and unified interface for application development.


\textit{\textbf{Comparison with other PLs:}}
The domains in which C/C++ library migrations occur differ significantly from those of other PLs. Unlike Java, JavaScript, and Python, where Testing, Network, and Serialization are the most frequently migrated areas~\cite{he_large-scale_2021, gu_self-admitted_2023}, the predominant domains for migrations in C/C++ are GUI and Build. 
The differences in migration domains can be attributed to their domain-specific applications. 
This shift also highlights the increasing significance and technical evolution in packaging and high-performance graphics rendering for C/C++ projects.
We can also see that the selection of migration alternatives for C/C++ developers tends to be consistent. One source library usually corresponds to no more than three target libraries (Fig.~\ref{fig:sub1}). 
While for Java developers, their choices vary a lot~\cite{he_large-scale_2021}. The details will be discussed in Sec.~\ref {sec:rq3}.

\vspace{-1mm}
\begin{result-rq}{Summary for RQ2:}
The domains with the most frequent library migrations in the C/C++ ecosystem are GUI and Build, underscoring both the complexity of the build process and the critical role C/C++ plays in graphics development.
Compared to other PLs (Java, JS, Python), C/C++ developers tend to migrate Build and application-oriented libraries (GUI) rather than basic libraries like testing, logging, and serializing. 
\end{result-rq}

\section{RQ3: How do developers choose their migration targets?}
\label{sec:rq3}

The migration targets represent the technology trend that developers switch to~\cite{DBLP:lib_selection, DBLP:lib_selection_2}, and their distribution reflects the divergence of technical trends in the C/C++ ecosystem. In RQ2, we identify different distributions and directions of migration targets in different domains. This question attempts to go further to characterize the migration alternatives in the C/C++ ecosystem.

\subsection{Method}
Library migration can be represented as a weighted directed graph where nodes represent libraries, edges represent migration rules, and the weight is the number of corresponding migration commits. 
Previous studies analyzed the differences in weighted degree (i.e., differences between weighted in-degree and out-degree) by $Flow(l)$ to demonstrate the extent to which a library is more likely to be adopted or abandoned~\cite{he_large-scale_2021}. However, the distribution of the weighted degrees (i.e., diversity of target libraries) is neglected, which is another important aspect of selecting targets. 
Inspired by the approach adopted by prior studies~\cite{he_large-scale_2021, gu_self-admitted_2023, python_library_migration_2024},
we extend the features of the migration graph to calculate both the distribution and difference of weighted degree, referred to as \textbf{diversity} and \textbf{directionality} respectively, 
to analyze the distribution of migration targets in C/C++.
\par \textbf{Diversity}
Choosing an appropriate target is crucial for library migration, especially from multiple candidates. 
The more candidates there are, the more challenging the selection may be.
We use $Entropy(l)$  to describe the diversity of candidates for a source library $l$ (Equation \ref{eq:entro}). Entropy is used to quantify the uncertainty in information. Here, a high entropy of a source library indicates that developers exhibit divergent perspectives on the alternatives for this library, and more importantly, the distribution of their choices tends to be rather uniform. If the entropy of a source library is zero, all developers choose the same target.

\vspace{-0.5mm}
    \begin{align}
    &Entropy(l) = -\sum\limits_{l_t \in \mathcal{L}} p(l, l_t)\log_2{p(l, l_t)} \label{eq:entro} \\
    &p(l_1, l_2) = \frac{|\{\langle c, l_1, l_2 \rangle\ |\ \exists c\in\mathcal{C}_m, \langle c, l_1,l_2\rangle\in\mathcal{M}\}|}{|\{\langle c, l_1, l \rangle\ |\ \exists c\in\mathcal{C}_m\wedge l\in\mathcal{L}, \langle c, l_1,l_2\rangle\in\mathcal{M}\}|}\notag
    \end{align}
\vspace{-0.5mm}

\par \textbf{Directionality}
Some libraries are frequently deprecated or adopted (e.g., \textit{jsonformoderncpp} and \textit{nlohmann\_json}). 
However, some libraries might be adopted by a subset of developers while being abandoned by others (e.g., \textit{gtest} and \textit{Catch2}). Understanding the adoption-to-abandonment rate of library migrations, i.e., the directionality of migrations, may inform us about which libraries have been phased out and which ones remain in a competitive state within the ecosystem.

Following previous studies~\cite{teyton2012mining, he_large-scale_2021}, we use $Flow(l)$ to describe the directionality of a source library $l$ (Equation~\ref{eq:flow}). Here we take the absolute value of the $Flow(l)$, focusing only on the magnitude of directionality rather than distinguishing whether libraries are adopted or abandoned. If $Flow(l)=1$, it indicates the library $l$ is always adopted or abandoned.

\begin{equation}
\vspace{-0.5mm}
\begin{aligned}
    Flow(l) = & \mid\frac{deg^-(l) - deg^+(l)}{deg^-(l) + deg^+(l)}\mid
\end{aligned}
    \label{eq:flow}
\vspace{-0.4mm}
\end{equation}

\subsection{Results}

\subsubsection{Diversity of target libraries}

\begin{figure}
    \centering
    \includegraphics[width=0.7\columnwidth]{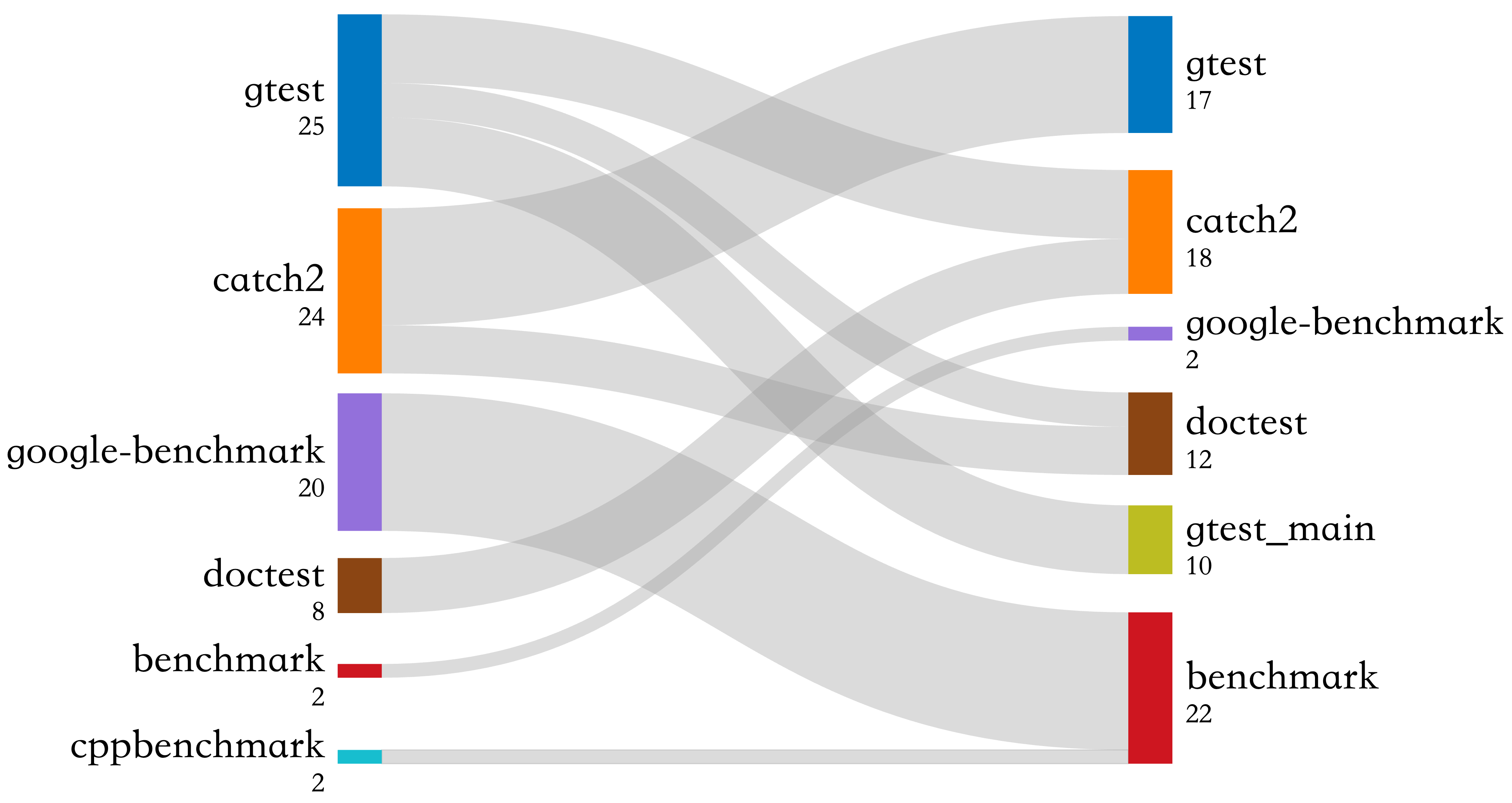}
    \vspace{-5mm}
    \caption{Migration subgraph for Testing libraries}
    \label{fig:test_sub}
    \vspace{-3mm}
\end{figure}

To illustrate entropy clearly, we visualize a sub-graph for a typical migration domain (Fig.~\ref{fig:test_sub}).
It shows a completely different migration situation from Fig.~\ref{fig:sub3}, where a source library corresponds to multiple target libraries and a library is abandoned in one migration while adopted in another. In this subgraph, $Entropy(gtest)=1.52$, $Entropy(catch2)=0.60$, and $Entropy(doctest)=0$. It indicates that the migration target of \textit{gtest} exhibits a higher degree of uncertainty, as it splits into three targets, with two having similar proportions, compared to \textit{catch2}, which migrates to two targets, one being superior to the other. While \textit{doctest} has only one migration target.

Furthermore, we calculate the entropy of all source libraries in C/C++.
Fig.~\ref{fig:entropy-dist} shows its extremely distorted distribution of $Entropy(l)$.
For 83.46\% of the source libraries (449~/~538), $Entropy(l)=0$, which means that developers only have one alternative option (i.e., one-to-one migration). 
For the remaining libraries with multiple targets, 17\% of them (16~/~89) have an entropy lower than 0.88 (calculated by $|\mathbf{L_t}|=2$ and $p(l,l_1)=0.7$).
It means that although there may be more than one alternative, at least 70\% migrations still correspond to the same target library.
Developers may have reached consensus or followed existing practices when conducting such migrations.

However, 44 libraries (8.18\%) have relatively high entropy (0.97 for $|\mathbf{L_t}|=2$ and 1.57 for $|\mathbf{L_t}|=3$), indicating a huge difference in target library selection. Among these libraries, GUI and Testing libraries needs attention. For example, for the Graphics library \textit{glew}, 11 migrations choose \textit{glad} to replace it. However, 8 choose \textit{epoxy} and others choose \textit{sld2}, \textit{glbinding}, and so on. (Testing libraries are shown in Fig.~\ref{fig:test_sub}) In this case, developers need to consider their requirements carefully and choose an appropriate alternative.  

\subsubsection{Directionality of the library migrations}

\begin{figure}[t]
    \centering
    \begin{minipage}[b]{0.49\columnwidth}
        \centering
        \includegraphics[width=\linewidth]{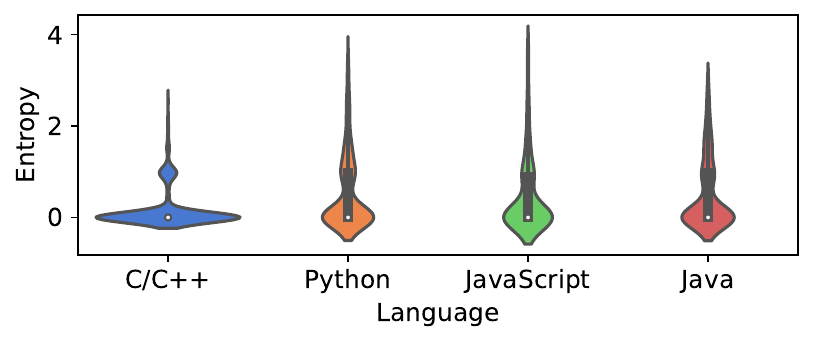}
        \caption{Distribution of domains for migrations in C/C++}
        \label{fig:entropy-dist}
    \end{minipage}
    \hfill
    \begin{minipage}[b]{0.49\columnwidth}
        \centering
        \includegraphics[width=\linewidth]{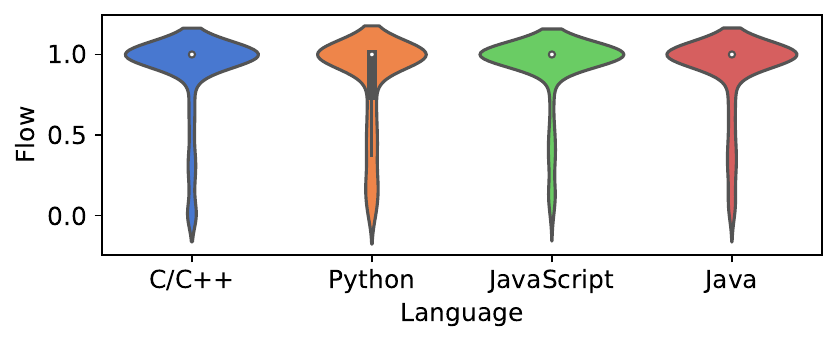}
        \caption{Migration subgraph for GUI libraries}
        \label{fig:flow}
    \end{minipage}
    \hfill
\end{figure}

Fig.~\ref{fig:flow} shows the distribution of $Flow(l)$ for source libraries $l$. 
Similar to $Entropy(l)$, the highly skewed distribution of migration flows indicates that the vast majority of libraries are always adopted or abandoned. 79.60\% of migration rules are entirely unidirectional, suggesting that the target libraries are superior to the ones they supplant in all aspects. 

\textit{\textbf{Comparison with other PLs:}}
Fig.~\ref{fig:entropy-dist} clearly shows that C/C++ exhibits a distinct tendency towards one-to-one migrations (83.46\%), in significant contrast with migrations observed in Python (63.32\%), JavaScript (69.57\%), and Java (65.28\%). This huge difference does not change obviously, even if we exclude libraries with migrations that occur fewer than five times. 
Factors inherent to the C/C++ ecosystem may contribute to this phenomenon. First, the C/C++ ecosystem has been characterized by well-established libraries such as \textit{boost} and \textit{qt}. These libraries are highly optimized and widely used, where developers may have formed a consensus. The same source may release different libraries for various requirements or architectures.
Secondly, the complexity in managing dependencies makes C/C++ developers conservative when choosing dependencies. They always follow established standards and best practices in communities.
Thirdly, the fragmentation of package management platforms may impose certain limitations on C/C++ developers when selecting software packages.
For the distribution of $Flow(l)$, JavaScript migrations (84.68\%) exhibit the highest degree of unidirectionally, while Python migrations (71.43\%) demonstrate the lowest. This may be the result of the characteristics of ecosystems. Libraries in the Python ecosystem face intense competition, whereas the JavaScript ecosystem exhibits a high rate of library deprecation~\cite{gu_self-admitted_2023}. 

\vspace{-1mm}
\begin{result-rq}{Summary for RQ3:}
Although some libraries have multiple
options in domains like GUI and Testing, for 83.46\% of C/C++ libraries, migrations are one-to-one. 79.60\% of migration rules are entirely unidirectional in C/C++. 
The much higher one-to-one and unidirectionality proportion compared to other languages suggests that C/C++ developers may have formed a consensus on migration. 
\end{result-rq}

\begin{table*}[t]
  \centering
  \caption{Definitions and distribution of migration rationales in C/C++. The rationales marked with † have extended definitions compared to previous study~\cite{he_large-scale_2021,gu_self-admitted_2023}, and the extended definitions are highlighted with \uline{underlines} and \textbf{bolded}.}
  \label{tab:reason}
  \renewcommand{\arraystretch}{1.4}
  \footnotesize
  \tabcolsep=0.3cm
  \label{tab:reason-def}
  \begin{tabular}{lp{11cm}r}
  \toprule
   \textbf{Rationale} &
    \textbf{Definition} &
    \textbf{Frequency} 
    \\ \midrule
    \textbf{Source Library} & Problems in the source library & 11 (16.18\%)\\
   \myindent Deprecation &
    The source library is inactive, not maintained, will be deprecated in the future, or not recommended to use officially. &

    3 (4.41\%)
    \\
   \myindent Bug or issue† &
    1)~Source library has bugs or emits warnings. 
    2)~\uline{The source library is \textbf{not stable} or \textbf{can not be compiled successfully} on certain platform.} &
    8 (11.8\%)
    \\
    \midrule
    \textbf{Target Library} & Advantages in the target library & 30 (44.12\%) \\
   \myindent Functionality &
    1)~The target library has new and powerful functionalities for the project to implement their desired features. 
    2)~The target library has more functionalities and makes the project more adaptable to future changes.  &
    14 (20.59\%) \\
   \myindent Usability &
    1)~The target library is easy to install, use, or maintain; the target library has better documentation. 
    2)~Using the target library can bring ease of implementation and cleaner code.  &
    4 (5.88\%)\\
   \myindent Performance† &
    1)~Using the target library can improve runtime performance. 
    2)~\uline{Using the target library results in \textbf{less compiling effort}.}  &
    5 (7.35\%)\\
   \myindent Activity &
    Although the source library is under maintenance, the project still chooses to use a more recent, well-maintained, or stable dependency. &
    5 (7.35\%)\\
   \myindent Popularity &
    The target library is more widely used or complies with industrial standards/ecosystem best practices. &
    1 (1.47\%)\\
   \myindent Size/Complexity &
    The target library is simpler or results in a smaller binary size. &
    1 (1.47\%)
    \\ \midrule
    \textbf{Project Specific} & Project specific requirements & 27 (39.71\%)\\
   \myindent Integration &
    1)~The project needs to integrate with different operating systems, platforms, or architectures. 
    2)~The project has to resolve conflicts between dependencies.  &
    15 (22.06\%)\\
   \myindent Simplification† &
    1)~The project migrates to reduce the number of dependencies. 
    2)~\uline{The project \textbf{no longer needs certain functionality} and replace relevant libraries.} 
    3)~\uline{The project migrates to \textbf{unify the package hosting platform}.} &
    12 (17.65\%)
    \\ \bottomrule
  \end{tabular}
  \end{table*}

\section{RQ4: Why do developers conduct C/C++ migrations?}
\label{sec:rq4}
\subsection{Method}
Reason implies the feature of migration from a deeper perspective. Understanding why developers conduct a migration can help them avoid possible migration and help related tools recommend proper alternatives.
SALMs identify migrations based on the clarification in commit messages. However, not all developers choose to explain why migrations are conducted. Given the large dataset in our work (7,083 migrations), it is extremely challenging to check each message with manual effort. 
Therefore, we first filter the commit messages by some keywords that are often used to describe migration reasons (e.g. because, for, so that, deprecate, better, can, help). Such keywords are summarized and refined from previous work~\cite{he_large-scale_2021, gu_self-admitted_2023}. After filtering, we get 316 records containing those migration-related keywords. One author first labels these records based on previous rationale framework~\cite{he_large-scale_2021, gu_self-admitted_2023} and generates a coding book for C/C++ migration rationales for the first iteration. Then, two authors independently label migration-related messages using this coding book. Since a message can involve multiple reasons, we use Krippendorff’s alpha with MASI distance to measure the inter-rater agreement~\cite{krippendorff2018content}. The Krippendorff’s alpha of the second iteration is 0.86, which is higher than the recommended score of 0.8~\cite{krippendorff2018content}. We finally got three themes and ten sub-themes. The detailed information is shown in Section~\ref{sec:rq4}.

It is worth noting that we also compare our findings in C/C++ with previous work in Python, JavaScript, and Java. For the diversity (part of RQ3) that has not been studied before, we calculated the results using publicly available datasets~\cite{he_large-scale_2021, gu_self-admitted_2023}.

\subsection{Results }
Table~\ref{tab:reason} shows the migration rationale in C/C++, including 10 categories among three themes (\textit{source library}, \textit{target library}, \textit{project-specific}) summarized from 68 migration commits with documented reasons.
Four new types 
are discovered in C/C++ compared to the framework proposed
in previous studies~\cite{he_large-scale_2021,gu_self-admitted_2023}, as highlighted with $\dag$ in Table~\ref{tab:reason}.
To avoid repetition, we omit explanations for rationales already covered in previous studies. The newly extended rationales are as follows:
\par{\textbf{Bug or issue under \textit{Source Library}:}} 
The source library is not stable or can not be compiled successfully on certain platforms.
C/C++ libraries are usually used in operating systems, where a variety of distributions necessitate library adjustments and compatibility measures. Therefore, libraries may not be stable or exhibit bugs on certain platforms. 
For example, \textit{\texttt{gcr-4} is not yet stable for gnome 43}~\cite{ExampleOfBug}. 
\par{\textbf{Performance under \textit{Target Library}:}}
Using the target library results in less compiling effort.
Compiling effort and time is a factor influencing developers' choices among libraries due to the lengthy compiling time of C/C++. For example, \textit{Removed unused \texttt{OpenSSL} dependency (replaced with \texttt{CryptoPP} due to static linking and compile times)}~\cite{ExampleOfPerformance}.
\par{\textbf{Simplification under \textit{Project Specific}:}}
The project migrates to unify the package hosting platform.
Given the variety of C/C++ PMTs, some developers try to change and unify their dependency management methods. For example, they might choose to only use libraries from Conan or Vcpkg, e.g., \textit{Use cmark for commonmark parsing (because \texttt{md4c} isn't in Vcpkg)}~\cite{ExampleOfSimplification}. This reason highlights the complexity of C/C++ dependency management. Developers try to alleviate this problem by adopting libraries from the same package hosting platforms. Another reason for simplification is that the project no longer needs certain functionality and replaces relevant libraries, e.g., \textit{Use plain \texttt{libcanberra} instead of \texttt{libcanberra-gtk3}. We no longer use the gtk-aware ca\_context, the dependency can be lowered now}~\cite{ExampleOfSimplification2}.

The third column of Table~\ref{tab:reason-def} presents the distribution of rationales. 
The complex development environment and cross-platform requirements of C/C++ projects result in a high proportion of migration due to project-specific reasons (39.71\%). Advantages in the target library (44.12\%) are another main reason, driven by the need for new features in certain domains(e.g., GUI) and performance requirements of C/C++.
  
\textit{\textbf{Comparison with other PLs:}}
When conducting C/C++ library migrations, developers tend to consider more advantages in the target library and project-specific requirements. This preference is aligned with Python~\cite{gu_self-admitted_2023}, but the reasons 
are distinct. 
While Python developers need to consider compatibility with both Python 2 and Python 3, C/C++ developers need to consider different OS distributions and architectures. 
Python is designed for easier use and development, so functionality and usability are favored by developers. In contrast, C/C++ development is often perceived as challenging, with a lengthy compiling time. 
Migrations in C/C++ usually have clear targets. 
Therefore, more functionalities, better usability, and less compiling effort have become goals pursued by C/C++ developers.
For example, simplify the size of dependency (e.g., $\langle \texttt{boost}, \texttt{boost-range} \rangle$).
Meanwhile, the proportion of simplification (17.65\%) in migration reasons is similar to Java (18.54\%)~\cite{gu_self-admitted_2023}.
This similarity can be attributed to the frequent use of these two languages in large-scale projects, where maintaining only essential features and compatible libraries is crucial. 
An intriguing observation 
is the C/C++ developers' effort to unify their package management methods through migrations. This indicates that various PMTs indeed pose significant challenges for developers, while consistent package management (e.g., Maven) can greatly simplify dependency tracking. 

\vspace{-1mm}
\begin{result-rq}{Summary for RQ4:}

C/C++ developers often migrate to leverage new features of the target library, integrate across various platforms, or simplify dependency management.
Four reasons not discovered by the prior studies are unveiled: stability, less compile time, abandonment of unnecessary functionality, and unification of dependency management.

\end{result-rq}

\section{Discussion}
\subsection{Takeaways}
\subsubsection{Evolution of package management tools} 

As discussed before, C/C++ dependency management is notoriously fragmented and complex~\cite{miranda2018use} compared to other languages. 
To manage the complexity, C/C++ evolves.
The number of migrations in emerging PMTs like Meson and Vcpkg is on the rise, diverging from the traditional tools that have been staples for many years.
While Deb's historical significance and widespread adoption have resulted in its highest number of migrations, the landscape of package management in C++ is evolving with the emergence of modern tools such as Meson and Vcpkg. 
Most migrations occurring in Deb are building and packaging-related tools. 
Modern package managers such as Vcpkg and Conan simplify the process of managing dependencies. They offer declarative and intuitive configuration files that streamline the addition, update, and removal of libraries. They are also able to support a wide range of platforms and compilers, and integrate smoothly with modern build systems (CMake) and continuous integration (CI) pipelines.
Tools like Meson offer advanced features, improved performance, and better usability compared to traditional methods~\cite{meson}. 
However, the migrations between different management tools also haunt developers.
We observe that a considerable number of projects adopt more than one tool to manage C/C++ dependencies, which leads to much maintenance cost.
Considering the huge effort of migrating PMTs from one to another, we suggest that evelopers choose a more modern tool to manage their dependencies.




\subsubsection{Tools for C/C++ migrations}
Researchers have dedicated significant efforts to automated tools for migration recommendation and code adaptation~\cite{DBLP:java_api_recommend, DBLP:java_api_mig_tool, DBLP:java_method_mig, DBLP:api_migration_pattern, DBLP:ase_api_migration, DBLP:icpc_migration_edit}.
Despite the fragmentation and complexity inherent in C/C++ package management, the highly unidirectional and one-to-one characteristics of library migrations in C/C++ help alleviate some of the burdens when developers select alternatives. Unlike other languages, C/C++ developers can significantly benefit from established best migration practices, reducing the complexity involved in selecting target libraries. One-to-one migration rules allow developers to focus on directly replacing old libraries with new ones that offer better performance, features, or compatibility.
To leverage these advantages, the focus of migration or recommendation tools for C/C++ could prioritize mining migration rules and automating code adaptation rather than evaluating which new library to adopt. Package hosting platforms could document best practices, serving as guidelines for similar migration needs. 
We also observe that some major updates (\Code{Qt}) even refactor the whole library, leading to one library being separated into different new components. Migration from such an update is not easy, and the qt-wiki has even provided corresponding documents~\cite{qtwiki}.
Therefore, such tools could support migrations between two collections of libraries. Developers in certain domains can also adopt modular design principles or leverage automated tools to mitigate the complexity of these migrations.

\subsubsection{Adoption of libraries}
The migration of libraries within C/C++ projects is often driven by three rationales (functionality, integration, and simplification) from two themes (target library and project-specific reasons) 
Cutting-edge features, intuitive APIs, extensive documentation, and faster compile times occupy a significant portion. While project-specific concerns vary, dependency conflict or license compatibility should be taken seriously for each project. 
Moreover, using libraries available on the same package management platform may represent a novel and beneficial approach to dependency management. 
The rationale for C/C++ migration indicates that developers could consider the specifics of projects (e.g., supported platform, performance requirements, and even PMTs) when adopting a library.

\subsection{Threats to Validity}

\subsubsection{Internal Validity}

Our study relies on parsing dependency profiles to identify C/C++ dependencies~\cite{tang2022}. However, an inherent threat 
lies in the fact that the listed dependencies might not be utilized. This discrepancy arises because dependency files typically enumerate all potential dependencies required for the code compilation but do not necessarily indicate which dependencies are active or relevant during runtime. Migrations from third-party libraries to standard libraries are also excluded because no new libraries are introduced.
We utilize static analysis to identify dependencies in C/C++-related configuration files. For some dynamically loaded dependencies, we may miss them or overestimate the dependencies.
We filter the reason by certain keywords, which means we may miss some rationales mentioned by developers with other keywords.
Manual efforts are also devoted to reason coding, which may introduce bias. Two authors independently code the migrations and discuss them to alleviate this threat. 

\subsubsection{External Validity}

Firstly, our analysis depends on migration rules identified by commit messages explicitly mentioning migrations. Migration rules not explicitly mentioned in commit messages might be overlooked, potentially leading to an incomplete representation of migration activities. 
However, we believe the automated identification of self-admitted migrations tends to have higher accuracy, and their commit messages contain richer information, providing deeper insights into the process and rationale behind such changes. 
The rule-matching approach allows us to detect and analyze migrations that follow the same rules even when they were not explicitly mentioned in the commit messages, thereby providing a more comprehensive view of migration practices. 
Secondly, our findings may not be generalizable to projects that use other package managers, build systems (e.g., CMake), or code clones. CMake scripts always involve intricate and dynamic scripting, making it difficult to reliably extract dependency information. Code cloning is also hard to be effectively and accurately detected. We choose as many dependency management methods as possible to mitigate this threat, including package managers and building systems, and analyze more than 120k blobs with dependency changes. And the way developers manage their dependencies should not significantly influence the characteristics of migrations we have found.
Finally, the generalizability of our results to programming languages other than C/C++ is limited. For different languages, different migration characteristics may occur due to distinct application domains and development practices. Therefore, we compare the migration patterns in C/C++ with other programming languages~\cite{gu_self-admitted_2023, he_large-scale_2021}. Both the divergences and similarities in how migrations occurred across different languages are revealed in our findings.

\section{Conclusion}

In this paper, we conduct a multidimensional comparative analysis of the prevalence, domain, target, and rationale of migrations in the C/C++ ecosystem and present a C/C++ migration knowledge base that offers valuable insights for researchers and developers. Our main contributions are:

1) A semi-automated mining method identifying migrations for C/C++ libraries based on git repositories.

2) The insight of dependency management in the C/C++ ecosystem and distinctive characteristics of C/C++ migrations and their differences compared to migration practices in Python, JavaScript, and Java.

3) First migration dataset for C/C++ libraries, which can serve as a foundation for further research and related tool development.

For future work, we plan to expand our investigation of C/C++ package migrations by incorporating CMake in a more comprehensive ecosystem analysis. 
Additionally, we identify several challenges in C/C++ ecosystem migrations that warrant further study.

\bibliographystyle{ACM-Reference-Format}
\bibliography{references}

\end{document}